\documentclass[preprint,12pt]{elsarticle}
 \usepackage{graphicx}
 \usepackage[matrix,frame,arrow]{xypic}
\input{Qcircuit}
 \usepackage{epsfig}
\usepackage{amssymb}
\usepackage{amsmath}
\begin{document}
\begin{frontmatter}
\title{Quantum Game Application to Spectrum Scarcity Problems}
\author{O.G. Zabaleta, J. P. Barrang\'{u}  and C.M. Arizmendi}
\address{Instituto de Investigaciones Cient\'ificas y Tecnol\'ogicas en Electr\'onica (ICYTE),\\
Facultad de Ingenier\'{\i}a,\\
Universidad Nacional de  Mar del Plata,\\
Av. J.B. Justo 4302, \\
7600 Mar del Plata, Argentina}
\begin{abstract}
Recent spectrum-sharing research has produced a strategy to address spectrum scarcity problems. This novel idea, named cognitive radio, considers that secondary users can opportunistically exploit spectrum holes left temporarily unused by primary users. This presents a competitive scenario among cognitive users, making it suitable for game theory treatment. In this work, we show that the spectrum-sharing benefits of cognitive radio can be increased by designing a medium access control based on quantum game theory. In this context, we propose a model to manage spectrum fairly and effectively, based on a multiple-users multiple-choice quantum minority game. By taking advantage of quantum entanglement and quantum interference, it is possible to reduce the probability of collision problems commonly associated with classic algorithms. Collision avoidance is an essential property for classic and quantum communications systems. In our model, two different scenarios are considered, to meet the requirements of different user strategies. The first considers sensor networks where the rational use of energy is a cornerstone; the second focuses on installations where the quality of service of the entire network is a priority.
\end{abstract}
\begin{keyword}
Quantum games \sep minority game \sep spectrum allocation


\end{keyword}
\end{frontmatter}

\section{Introduction}
Modern wireless communications networks are composed of users accessing the network through multiple devices, including cellular phones, Wi-Fi devices, and GPS receivers; moreover, users often operate multiple applications simultaneously. The widespread use of these devices demands heavy use of network resources. The number of wireless devices and applications has grown exponentially in recent years, creating an almost unfathomable radio spectrum demand. Radio spectrum assignments are static and mainly assigned to services such as TV and radio broadcasts, navigation, and so forth. As a consequence, few spectra are unused, making it a scarce and extremely valuable resource. Nevertheless, numerous studies have found that licensed spectrum is considerably underutilized in temporal, spatial, and frequency domains \cite{book:ChunSheng2015}. By considering spectrum scarcity problems caused by static spectrum allocation, cognitive radio (CR) is viewed as a novel approach for improving the utilization of such an important resource \cite{Haykin2005}. The main idea of CR is that users without licenses (cognitive users) can sense the spectrum in order to detect the presence or absence of licensed users (primary users); this enables them to access licensed frequency bands when primary users are not present.
Thus, in a framework of spectral opportunities, the secondary users must be able to make decisions and negotiate in the short term. By thinking of the cognitive users as players competing or cooperating to access available resources, the outlined scenario can be modelled by means of game theory.

\subsection{Quantum games and Communications}\label{sec:over}
Game theory is a mathematical tool that analyses the strategic interactions among multiple decision makers. The generality of the theory permits its use for modelling a wide variety of problems from different research areas \cite{Osborne}. The design of fair, secure, and efficient quantum information protocols is necessary to guarantee the development of reliable quantum networks. Resource allocation is one of the most important stages, and can be viewed as a competition in which the players are the nodes in a network that can control the nodes actions. Furthermore, several authors have tackled the design of transmission protocols on classical networks by using game theory techniques, and have obtained interesting outcomes \cite{Wang2010,Zhen2010}. We recently applied quantum games to quantum wireless networks \cite{DublinABZ,Zaba2014}, in order to enhance their efficiency.\\
Quantum games have proven useful for solving problems encountered in the decision sciences, in which the most relevant case is the prisoner's dilemma. In the original classical version, the Nash equilibrium represents an inconvenient situation for both players, while in the quantum prisoners game, a new Nash equilibrium appears that is both Pareto optimal and a better situation for both players \cite{Eisert1999}. Moreover, for certain games, quantum strategies have proved to be more effective than classic strategies. This is the case for the classical penny-matching game presented by Meyer, \cite{Meyer1999} and the battle of the sexes game considered by Marinato et al., who showed that the introduction of entangled strategies leads to a unique solution, whereas in the classical case, the theory cannot make any unique prediction \cite{Marinato2000}.
Furthermore, in cites \cite{Zaba2010,ArizZaba2011}, a quantum formulation of the dating market problem was introduced. In \cite{Schmidt2013}, the authors quantize the gamble known as Russian roulette. More recently, \cite{Pawela2016179} studied the advantages of quantum strategies in evolutionary social dilemmas on evolving random networks, focusing on two-player games such as the prisoner's dilemma, snowdrift, and stag hunt. Quantum game theory has been applied to a wide variety of phenomena where quantum laws rule; these include social decision theory \cite{Yukalov2015}, bioprocesses that obey quantum statistical mechanics \cite{Clark2012,Clark2010}, and quantum communications.\\

In previous works, we analysed quantum MAC algorithms that use quantum game tools as a method of providing fair access to network users. In \cite{rpic2013}, a quantum algorithm was proposed to improve the current classic protocols. There, under a hybrid cellular wireless network, users are capable of communicating in a centralized manner. Specifically, they communicate under the control of a base station (BS), a device in charge of receiving and transmitting signals to mobile devices in the network \cite{ChunSheng2015}, or (eventually) in a distributed mode where communication among users is direct (i.e. without the intervention of a third party). In the second mode, the network nodes behave as non-cooperative game players that must decide the moment to transmit by analysing the other players actions. The channels limited capacity and the multiple users wanting to transmit shape a scenario in which channel performance improves when fewer users are attempting to transmit in the same time slot. The minority game has been widely used in situations in which players compete for limited resources, such as choosing which evening to visit a bar that is usually overcrowded. In this context, we propose a model to fairly and effectively manage the resource allocations in cognitive radio networks, based on a multiple-users multiple-choice quantum minority game \cite{DU2002,Chen2004}.

The paper is organized as follows. Section~\ref{sec:Twousers} contains a description of the model in the form of a two-user system, which acts as a preview of the more general case further detailed in Section~\ref{sec:Nusers}. In Section \ref{sec:CirDesc}, a quantum circuit is proposed to implement entanglement. In Section~\ref{sec:altern} two different alternatives and their implications are described. Finally, conclusions and further work are depicted in Section~\ref{sec:Conc}.

\section{System model}\label{sec:Twousers}
The system analysed in this work has $N$ channels and $N$ users that must be assigned to one of those channels. The state of such a system in Dirac notation of some user $j$, where $j=0,1,...,N-1$, is $\ket{c_j}$, with $c_j = 0,1,...,N-1$. Moreover, the state of the entire system $\ket{\psi}=\ket{c_0} \otimes \ket{c_1}\otimes ...\otimes\ket{c_{N-1}}=\ket{c_0 c_1...c_{N-1}}$. Thus, it must be understood that user $0$ is assigned to channel $c_0$, user $1$ is assigned to channel $c_1$, and so on. In order to facilitate understanding, we present the simplest case of two users and two channels.

\subsection{Two-user game}

Let $0$ and $1$ be the indexes of two smart devices attempting to transmit information through two free channels, $0$ and $1$. The devices are assumed to be indistinguishable, and thus have identical transmission preferences. The states of the system are represented by vectors of a Hilbert space; more specifically, the vector position corresponds to the user, and the value in each position represents the user's assigned channel. Collisions are avoided when channels are not shared. For example, a desirable state is $\ket{c_0 c_1}=\ket{10}$, which specifies that user $0$ is assigned to channel $c_0=1$; meanwhile, user $1$ is assigned to channel $c_1=0$. If players play classically, the probabilities of each user and channel are all equal. The quantum equivalent for that case is $$\ket{\psi_C}=\frac{(\ket{00}+\ket{01}+\ket{10}+\ket{11})}{2},$$. Then, according to the classic strategies, it is clear that they have (at best) a 50/50 chance of avoiding collisions, and no strategy can modify the system in order to avoid collisions completely. Therefore, it is a classic Nash equilibrium of the system \cite{Osborne}.
On the other hand, they can do better if they play quantum, because they can achieve a $100\%$ probability of success. In order to take advantage of quantum computing, a one-shot quantum game is proposed; it begins with the system in an entangled state, $$\ket{\psi_e}=\frac{(\ket{00}-\ket{11})}{\sqrt{2}},$$ which is a linear combination of two of the four possible states. In this manner, the state $\ket{00}$ means that both users are assigned to channel $0$; meanwhile, state $\ket{11}$ assigns both users to channel $1$. In order to change the initial state, the players must apply a strategy which, mathematically, is represented by a two-dimensional operator that we call $U$. Generally, players can choose to operate on their qubits using a classic or quantum $U$, in order to increment their chances of winning. However, there is only one optimal quantum strategy (the Hadamard gate $U=H$) that modifies the system to a more favourable state for the two-user example. Given the condition that is applied by both players, the final state $\ket{\psi_f}$ is:
\begin{equation}
\ket{\psi_f}=H^{\otimes 2}\cdot\frac{(\ket{00}-\ket{11})}{\sqrt{2}}=\frac{(\ket{01}+\ket{10})}{\sqrt{2}}
\end{equation}
From $\ket{\psi_f}$, it arises that the system can only collapse to $ (\ket{01}) $, where user $0$ is assigned to channel $0$ and user $1$ is assigned to channel $1$, or to $ (\ket{10}) $, where user $0$ is assigned to channel $1$ and user $1$ is assigned to channel $0$. Thus, by quantum rules, a new Nash equilibrium arises, where the worst case is avoided and both users transmit successfully with probability 1. Furthermore, it is a Pareto optimal solution because it is impossible to make any player better off without harming some other player.\\
It is important to note that, because the studied network is composed of indistinguishable devices, the necessary condition that all players take the same actions is natural.

\section{N-users game description}\label{sec:Nusers}
Usually, there are $N>2$ users sharing a spectrum assumed to be divided into $N$ channels. Because none of the users has information about other users, there is a high probability that more than one of them will take part in a collision. When that occurs, all those involved cannot transmit, resulting in a situation that must be avoided or, at least, minimized by means of appropriate spectrum allocation protocols. These types of problems are difficult to solve classically as the number of players increases, that is, they are included in the group referred to as NP problems \cite{Garey1979}. Accordingly, they cannot be solved in polynomial time, which generally results in inefficient solutions. We are facing a type of decision problem consisting of agents with similar objectives that compete for a limited number of resources. Therefore, the spectrum allocation problem may be modelled as a multiple-options minority game. \\
The CR concept implies that cognitive devices can make smart choices and access the spectrum holes left unused by primary users. Despite this promising idea, it is very important to take into account that the existence of those spectral holes is dynamic in size and limited in time, which causes difficulties in sensing, sharing, and allocating tasks. The first constraint determines how many users can transmit simultaneously. Meanwhile, the second constraint limits the time that users have to select the channel they will transmit in, and the time they have to transmit. As the number of CR users increases, the decision processes become more complex, thus limiting the time the users have to transmit. Taking the latter into account, an efficient spectrum allocation algorithm is absolutely necessary.\\
Many researchers, including us, point to the use of game theory as the most appropriate technique to model (and consequently perform) resource sharing and allocation tasks \cite{Xiao2005,Charilas2010,Pillai2014,Zaba2014}. In this same line of thought, we go a step further by proposing the use of a one-shot quantum game to minimize decision times and the number of collisions. \\

\begin{figure}[h!]
 \centering
  \includegraphics[width=13cm]{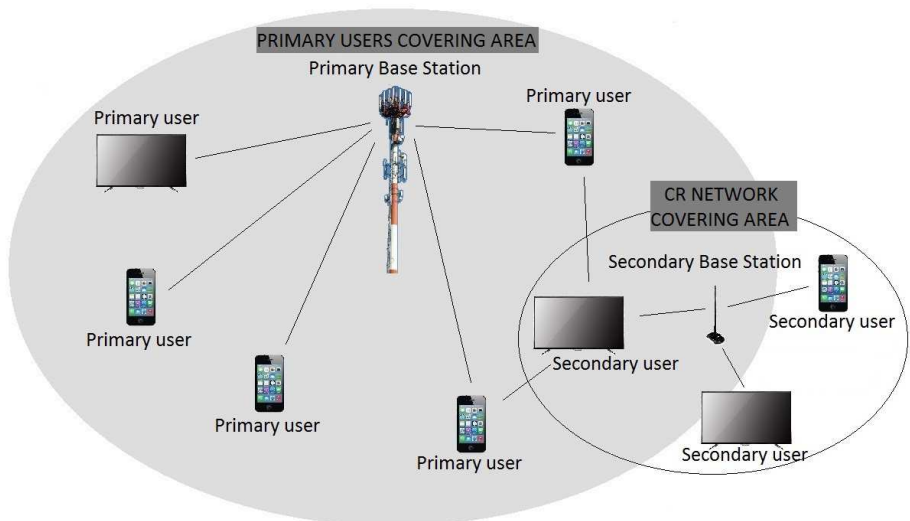}\\
  \caption{Cellular Cognitive Radio scheme}\label{fig:CR1}
\end{figure}
Our model considers a cellular network in which each cell has a single cognitive BS and a group of CR users in its coverage range. The qualitative scheme of the network is shown in fig. \ref{fig:CR1}. The BSs are transceivers in charge of connecting the devices to other devices in the cell. To achieve this, they collect the CR user reports, and prepare to allocate the radio channels. It is assumed that the devices cooperatively sense the spectrum and record information about the spectrum holes, which will eventually be provided to the base stations. Cooperative sensing has been previously analysed by other authors \cite{Mishra2006}. \\
In the following, we focus on a quantum algorithm capable of managing the spectrum allocation based on probability amplitude amplification. More specifically, we present two cases of interest: the first one aims to avoid all users being assigned to the same channel, and the second one aims to enhance the probability of quantum states that assign different channels to users. \\
The proposed quantum medium allocation evolves by following three basic steps:

\begin{enumerate}
\item The cognitive quantum BS assigns a set of qubits to the cognitive users in the cell range and prepares entangled state $\ket{\psi_e}$.
\begin{equation}\label{eq:ent}
\ket{\psi_e}=\frac{1}{\sqrt{N}}\sum_{k=0}^{N-1}\omega_{N}^{k\cdot p}\ket{kk\cdots k},
\end{equation}
where $ \omega_N=e^{2\pi i/N} $ and $p$ is a tunable parameter that modifies the amplitude phase. Depending on $p$, it is possible to select BS preferences to avoid the least favourable case, $p=1$, or, on the other hand, to enhance the optimum one, $p=\frac{(N(N-1))}{2}$.
\item Every node locally applies a one-shot strategy $U$ to the initial state, which makes the system collapse to a new state.

\begin{equation}
\ket{\psi_f}= U^N\cdot \ket{\psi_e},
\end{equation}
\item The nodes of each cell measure the final state $\psi_f$ to obtain the assigned channel.
\end{enumerate}
In what follows, we present the quantum circuits for the case N=4 and describe the main steps.

\section{Quantum circuit description}\label{sec:CirDesc}
\begin{figure}[h!]
\[
\Qcircuit @C=0.7em @R=0.5em {
&&\lstick{\ket{0}} & \qw & \multigate{1}{R^{\otimes 2}} & \qw & \ctrlo{1} 		& \qw & \ctrl{1} 		  	& \qw & \ctrl{1} 		  	 & \qw & \qw & \\
&&\lstick{\ket{0}} & \qw & \ghost{R^{\otimes 2}} 	  	& \qw & \ctrl{1} 			& \qw & \ctrlo{1} 		  	& \qw & \ctrl{1} 		  	 & \qw & \qw & \\
&&\lstick{\ket{0}} & \qw & \qw 				  			& \qw & \multigate{5}{F_1}& \qw & \multigate{5}{F_2}	& \qw & \multigate{5}{F_3}	& \qw & \qw & \\
&&\lstick{\ket{0}} & \qw & \qw 				  			& \qw & \ghost{F_1} 		& \qw & \ghost{F_2} 		& \qw & \ghost{F_3} 		 & \qw & \qw \gategroup{1}{12}{8}{13}{1em}{\}} & \\
&&\lstick{\ket{0}} & \qw & \qw 				  			& \qw & \ghost{F_1} 		& \qw & \ghost{F_2} 		& \qw & \ghost{F_3} 		 & \qw & \qw & & & \ustick{\ket{\psi_e}} \\
&&\lstick{\ket{0}} & \qw & \qw 				  			& \qw & \ghost{F_1} 		& \qw & \ghost{F_2} 		& \qw & \ghost{F_3} 		 & \qw & \qw & \\
&&\lstick{\ket{0}} & \qw & \qw 				  			& \qw & \ghost{F_1} 		& \qw & \ghost{F_2} 		& \qw & \ghost{F_3} 		 & \qw & \qw & \\
&&\lstick{\ket{0}} & \qw & \qw 				  			& \qw & \ghost{F_1} 		& \qw & \ghost{F_2} 		& \qw & \ghost{F_3} 		 & \qw & \qw & \\
&&\\
&&		&&	&	   & \lstick{\uparrow ~~} \\
&&\\
&&		&&	&	   & \lstick{\ket{\psi_1}} \\
}
\]
\caption{Circuit that generates the initial entangled state $ \ket{\psi_e} $.}\label{fig:entGen}
\end{figure}
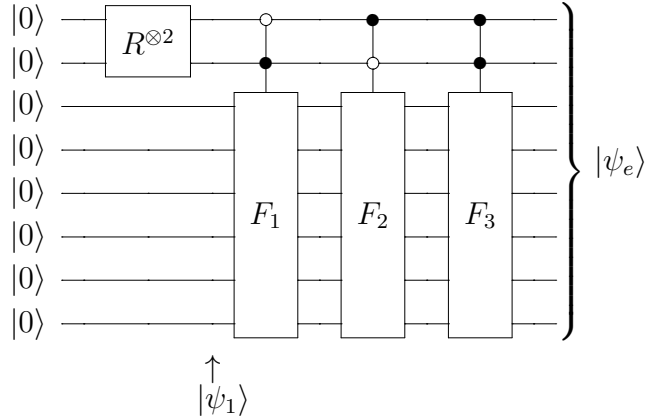

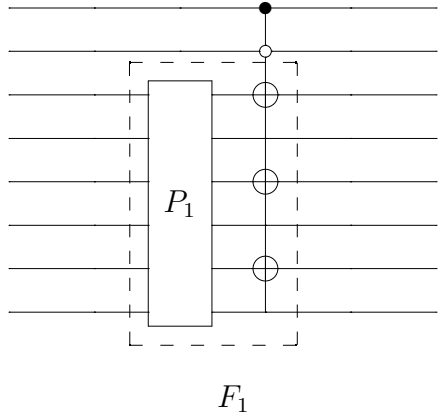
\begin{figure}[h!]
\[
\Qcircuit @C=0.7em @R=0.5em @!R @!C{
&& \qw & \qw 				& \ctrl{1} 	& \qw & \qw &\\
&& \qw & \qw 				& \ctrlo{6}	& \qw & \qw &\\
&& \qw & \multigate{5}{P_1}	& \targ		& \qw \gategroup{3}{4}{8}{5}{1.2em}{--} & \qw &\\
&& \qw & \ghost{P_1} 		& \qw 		& \qw & \qw &\\
&& \qw & \ghost{P_1} 		& \targ		& \qw & \qw &\\
&& \qw & \ghost{P_1} 		& \qw	 	& \qw & \qw &\\
&& \qw & \ghost{P_1} 		& \targ		& \qw & \qw &\\
&& \qw & \ghost{P_1} 		& \qw		& \qw & \qw &\\
&&\\
&&	   & 					& \lstick{F_1} &	  & 	&\\
}
\]
\caption{$Ctrl-F_1$ gate circuit, where $P_1=\omega_{N}^{1p}\cdot I^{\otimes 6}$. Looking from top to bottom, the $F_1$ operation is performed on the last six lines only if the state of the first two upper lines is $\ket{10}$.}\label{fig:Fkgate}
\end{figure}

Figure \ref{fig:entGen} shows a possible circuit to generate entangled state $\ket{\psi_e}$. The system in base state $\ket{00...0}$ is modified by the action of gate $R$ applied on the two upper qubits,
\begin{displaymath}
R=\frac{1}{\sqrt{2}}\begin{bmatrix}
~1 & 1 \\
-1 & 1 \\
\end{bmatrix}
\end{displaymath}
generating state $\ket{\psi_1}=\frac{\ket{00000000}}{2}-\frac{\ket{01000000}}{2}-\frac{\ket{10000000}}{2}+\frac{\ket{11000000}}{2}$.
Then, the two upper qubits of $\ket{\psi_1}$ are the control lines of three $Ctrl-F$ gates. A white circle in a control line indicates that the control qubit must be in state $0$; meanwhile, a black circle implies that the control must be in state $1$ in order for $F_k$ to be applied (see figure \ref{fig:Fkgate}). Note that the range of the system state is $N\cdot log_2(N)$ and that $R^{log_2(N)}$ must perform the rotation on the upper $log_2(N)$ qubits in the more general case. The extension of the circuit to $N$ is straightforward.

Finally, the action of gates $ F_k $ on state $\ket{\psi_1}$ yields:
\begin{equation}
\ket{\psi_e}=\frac{\ket{00000000}}{2}-\frac{\ket{01010101}}{2}-\frac{\ket{10101010}}{2}+\frac{\ket{11111111}}{2}
\end{equation}

\section{Two alternatives - distinct purposes}\label{sec:altern}
One of the main functions of cognitive radio is to provide a fair spectrum scheduling method among coexisting cognitive users. In this context, the purpose of the proposed methods is to improve the classic methods ability to decrease the collision probability. The objective of the first method is to increase the probability of occurrence of the no-collision state. When one user cannot transmit because of a collision, he must wait a lapse of time to re-manage the transmission request. System reliability and the networks quality of service improve if collisions are avoided. The second method is focused on wireless sensor networks, where the importance of all nodes being able to transmit is superseded by the importance of avoiding network downtime; here, the objective is to avoid the massive collisions that occur when all users are assigned to the same channel. \\
As was described in Section \ref{sec:Nusers}, the channel assignation procedure is the same in both cases. The base station prepares the entangled state of eq. \ref{eq:ent} with all the users in the cell, setting $p=N(N-1)/2$. After that, the users are positioned to perform their strategies. Strategy $U$ that applies each player is represented by an $N\times N$ unitary matrix whose elements are
\begin{displaymath}
U_w=\frac{1}{\sqrt{N}}(e^{2\pi i/N})^{r\cdot c},
\end{displaymath}
where $r,c=0,1...N-1$ are the row and column indexes.

Then, the final state is $$\ket{\psi_f}=U^{\otimes N}\ket{\psi_e}=  \left( \frac{1}{\sqrt{N}} \right)^{N+1}\sum _{ k=0 }^{ N-1 }{ \omega_{N}^{k\cdot p}\ket{kk\cdots k}},$$ $$\ket{\psi_f}= \left( \frac{1}{\sqrt{N}} \right)^{N+1}\sum _{ k=0 }^{ N-1 }\sum _{ c_0=0 }^{ N-1 } \dots \sum _{ c_{N-1}=0 }^{ N-1 } \left( e^{\frac{2\pi i}{N} k\cdot p} e^{\frac{2\pi i}{N} k\cdot c_0} \cdots e^{ \frac{2\pi i}{N} k\cdot c_{N-1}} \ket{c_0 \cdots c_{N-1}} \right).$$
Thus, the state coefficients can be expressed as:

\begin{equation} \label{eq:Coef}
\alpha_{c_0 \cdots c_{N-1}}= \left( \frac{1}{\sqrt{N}} \right)^{N+1}\sum _{ k=0 }^{N-1} e^{\frac{2\pi i}{N} k(\overbrace{p+c_0+c_1+ \cdots + c_{N-1}}^m) }.
\end{equation}

\subsection{Strategy to increase the optimum case probability}
The fairness of the network implies that every user has a priori the same chances to transmit. In the language of games, the BS acts as the arbiter of the game because it assigns the qubits to the players and creates the entangled state. Later on, the players strategies modify the state amplitudes and hence their chances to win. The players receive a reward, which in this case is to succeed in transmitting.\\
As set forth above, once spectrum holes are detected, the nodes must be assigned to one channel. Clearly, there are $N!$ possibilities that every player will be assigned to different states, with $N$ being the number of cognitive users. Therefore, provided that all the cognitive users are indistinct, the probability that all of them transmit at the same time is $P_c=\dfrac{N!}{N^N}$ in the classic world; for example, $P_c=2.4\times10^{-3}$ if $N=8,$. Such a low success probability can only be increased by means of statistical methods involving exploration and/or a previous knowledge of the network \cite{Sharma2015}, which is hardly possible if the network is continuously changing. In this framework, we propose the one-shot quantum game-based algorithm.\\
The $m$ sum in the phase factor of eq. \ref{eq:Coef} is analysed in order to properly select $p$. Thereby, a proper use of quantum interference makes it possible to improve the players chances.
The case where $c_0\neq c_1 \neq \cdots \neq c_{N-1}$ leads to $m=p+\frac{N(N-1)}{2}$. Thus, in order to guarantee the constructive interference, $p=\frac{N(N-1)}{2}$ and the phase factor is $e^{i2\pi k(N-1)}.$
Finally, the probability of the most favourable case is $P_{best}=N\cdot\dfrac{N!}{N^N}$, which is $N$ times larger than the classic one, $P_{best}=N\cdot P_c.$ Clearly, the algorithm performance provides a more efficient use of the devices energy, extending the time of communication and battery life. However, this point is even more sensitive in the type of networks that are analysed below.

\subsection{Strategy to avoid the most unfavourable situation}
Wireless sensor networks are a type of autonomous communication network mainly deployed in areas where access is almost impossible. Every device installed at each node is a small computer in charge of monitoring physical and environmental conditions such as temperature and air pressure. The sensed data are sent to a base station for analysis. Although sensor networks were originally designed for military purposes, their applications now include area sensing, industrial monitoring, and health care monitoring \cite{Boonsongsrikul2013,Yick2008}. One of the main challenges that communications engineers must face is the optimization of the networks power consumption, because of the limited lifetime of the devices batteries and the impracticality of replacing them frequently. Therefore, in order to extend the networks lifetime, more efficient communication protocols are needed. Because collisions are the main cause of unnecessary energy consumption, we propose a quantum algorithm that prevents the most unfavourable situation. When all users are assigned to the same channel, no transmissions will be performed. There are $N$ of these worst-case possibilities from a total of $N^N$, so the probability of the worst-case is $P_{worse}=N^{(1-N)}$ by means of classic computation, where the probability that any user will be assigned to any channel follows a uniform distribution. \\
Once the eventually free channels are identified, the channel allocation procedure begins. The proposed quantum algorithm considers that the cognitive BSs have the extra ability to share a set of qubits with each node in the cell and to prepare the entangled state of eq. \ref{eq:ent}, setting $p$ so that the probability amplitude associated with states $\ket{c,c,...,c}$ is reduced to zero. If players measure their state directly on $\ket{\psi_e}$, it will collapse to one of the worst cases. Otherwise, the users perform their strategies in order to change their chances.
Let us note that $c_0= c_1 = \cdots = c_{N-1}= c$ leads to $m=p+N\cdot c$ when all users apply $U_w$ as before. Thus, in order to guarantee the destructive interference, $p=1$ can be chosen. Then, the probability amplitude coefficients are:
 $$\alpha_{c c \cdots c}= \left( \frac{1}{\sqrt{N}} \right)^{N+1} \sum _{ k=0 }^{N-1} e^{i2\pi k c}e^{\frac{2\pi i}{N} k}=0.$$
By allowing at least one node to send information at a certain time slot by using the idle spectrum holes, the sensor network avoids downtime, a significant aspect regardless of the network structure \cite{Gurwinder2012}.
For instance, if the network uses a star topology, every node communicates directly with the BS. Because the nodes can communicate only through the BS, it represents a single point of failure (SPF) that makes this topology unreliable. However, owing to its simplicity, it is frequently chosen when coverage areas are not too wide. In that case, the quantum BS must prepare the new allocation scheme by requesting information from the rest of the CR nodes. Although the SPF problem remains unsolved because there are many failure sources, the network reliability is improved under normal BS functioning owing to the one-shot characteristic of the algorithm, which allows at least one node of the star to always send information; this optimizes energy use.
Meanwhile, in the case of multihop systems, each node can communicate directly and is able to take distinct paths to reach the data collector, which is advantageous as there is no SPF. On the other hand, these networks have an important disadvantage  high power consumption. To operate, they must draw more power because each node in a mesh must act as a BS. This issue is even more serious if the spectrum allocation task is not performed efficiently. Likewise, in our model, each node of the mesh must eventually prepare the allocation scheme following the procedure explained above, in order to exchange information gathered from the environment or from other nodes. The goal is to minimize the power consumption of mesh topologies by reducing collisions, which is made possible through the proposed quantum allocation algorithm.

\section{Conclusions}\label{sec:Conc}
In the context of spectral opportunities that constantly change with time, designing a fast and efficient method of spectrum allocation is increasingly necessary. In order to address the problem of many users competing for a common resource, (in this case, spectrum), many researchers have applied strategies based on game theory. The classic algorithms that have been attempted up to now need exploration and learning time to allow players to select the most favourable actions; this costs valuable time that can be used to transmit, and provides no guarantee of success. In the present work, we have proposed a quantum game-based scheme for cognitive spectrum allocation. The model offers two alternatives. The first aims to increase the no-collision probability over that of the classic approaches, which is essential in networks (such as cell phone networks) where quality of service is prioritized. The other alternative prevents all the cognitive users from converging to the same channel. This strategy is proposed for sensor networks, where one of the main requirements is that they never stop working; this allows the base stations to continuously receive data for analysing. Because of the characteristics of the one-shot algorithm, less time is wasted in the channel allocation process, which makes it possible to repeat the algorithm and further increase the success probability. Finally, both alternatives contribute to energy savings through the reduction in channel allocation times, an item that is even more sensitive in the case of sensor networks. In such a case, we considered two actual network topologies in which the proposed allocation algorithm can be successfully applied. Future trends in wireless sensor networks will impose more autonomy and less power consumption. The work we have presented takes advantage of cognitive radio and quantum game techniques to address these issues more efficiently, compared to the classic methods. Although cognitive radio and quantum communications are still in development, we believe our proposal advances the adaptation of these new communication paradigms.

\bibliographystyle{unsrt}

\bibliography{biblio}

\end{document}